\documentclass[pra, twocolumn, showpacs, superscriptaddress]{revtex4}
\usepackage{amssymb}
\usepackage{amsmath}
\usepackage{graphicx}
\usepackage{dcolumn}
\usepackage{color}
\usepackage{bm}
\usepackage{subfigure}
\usepackage{amsfonts}

\newcommand{\e}{\varepsilon}
\newcommand{\W}{\Omega}
\newcommand{\w}{\omega}
\newcommand{\Ph}{\Phi_{\Omega}}
\newcommand{\Ga}{\Gamma_{\Omega}}
\newcommand{\kapa}{\beta}

\begin{document}

\title{Classical and quantum dispersion-free coherent propagation\\ by tailoring multi-modal coupling}

\author{Aikaterini Mandilara}
\affiliation{Department of Physics, School of Science and Technology, Nazarbayev University, 53 Kabanbay Batyr Avenue, Astana, 010000, Kazakhstan.}
\author{Constantinos Valagiannopoulos}
\affiliation{Department of Physics, School of Science and Technology, Nazarbayev University, 53 Kabanbay Batyr Avenue, Astana, 010000, Kazakhstan.}
\author{Vladimir M. Akulin}
\affiliation{Laboratoire Aim\'{e}-Cotton CNRS UMR 9188, L'Universit\'{e} Paris-Sud et L'\'{E}cole Normale Sup\'{e}rieure de Cachan, B\^{a}t. 505, Campus d'Orsay, 91405 Orsay Cedex, France}
\affiliation{Institute for Information Transmission Problems of the Russian Academy of Science, 19 Bolshoy Karetny Per, Moscow, 127994, Russia.}

\begin{abstract}
It is shown that tailored breaking of the translational symmetry through weak scattering in waveguides and optical fibers can control chromatic dispersions of the individual modes at any order; thereby, it overcomes the problem of coherent classical and quantum signal transmission at long distances. The methodology is based on previously developed quantum control techniques and gives an analytic solution in ideal scattering conditions;  it has been also extended to incorporate and correct non-unitary effects in the presence of weak back-scattering. In practice, it requires scatterers able to couple different modes and carefully designed dispersion laws giving a null average quadratic dispersion in the spectral vicinity of the operational frequency.
\end{abstract}

\pacs{03.67.Hk  Quantum communication, 42.50.-p  Quantum optics, 42.79.Sz  Optical communication systems, multiplexers, and demultiplexers , 42.81.-i  Fiber optics}
\maketitle

\section{Introductory Comments \label{S0}}
One of the major hindrances in information transmission at long distances via multimode optical fibers, is the spread in arrival times of packages due to the discrepancy in the group velocities for various modes \cite{Book, Book2}. At the same time, the dispersion effects need to be compensated for each of the modes separately \cite{Roadmap}. In classical single-mode optical networks the most common solution is the dispersion compensating fibers. In the multimodal scenario, the techniques are based on the inter-conversion of the signal between different modes \cite{HighModes} or alternative elegant suggestions involving principal orthogonal modes \cite{PrincMode} and adaptive optical modulators \cite{MultiDisp}. As far as dispersion compensation in single-mode quantum signals containing low number of photons is concerned, the direct extension of classical methods might not be straightforward \cite{Gisin} since a compensation method may result in additional photon losses. Quantum solitons \cite{qsoliton3, qsoliton} potentially remedy such an issue but require sophisticated means of preparation.

This work elaborates the topic of overcoming these snags by emulating a multimode waveguide with identical group velocities of different modes and no quadratic dispersion. In particular, we show that breaking the translation invariance both compensates  for quadratic dispersion at any predetermined order $k$ of Taylor expansion and cancels the discrepancy in group velocities. To the best of our knowledge, no scheme has been suggested before solving this intriguing problem. The proposed technique  is based on previously developed ideas for implementing quantum control \cite{Vlad} over compact semigroups \cite{FE}   but it also goes  one step further, by incorporating and treating weak non-unitary effects (up to  second order) arising in the non-ideal scenario.

The suggested method provides  dispersionless propagation along the fiber/waveguide by perturbing the translational symmetry similarly to fiber Bragg gratings \cite{FiberGratings, Brag} and photonic band structures \cite{PBG, PBG2}. This becomes feasible via scatterers or mode couplers like those used in the design of dispersion compensation filters for two \cite{TwoModes, TwoModes2, TwoModes3} and higher number of modes \cite{HighModes}.The followed approach also bears some resemblance to multimode interferometers \cite{Sold1995}, where several propagating modes are constructively or destructively aggregated to formulate desired wavefront patterns. Importantly, there are also similarities with Talbot effect \cite{Talbot}, namely the periodically repeated self-imaging in a diffraction grating; indeed, the described concept concerns a link between two distant points (emitter, receiver) and thus repeats inevitably its unit cell and its, theoretically, dispersion-free response. The main contribution of our method is that the discrepancy in the mode dispersion laws can be compensated up to any order $k$ of the Taylor expansion around a given operational signal frequency, in complete analogy with the regime of quantum error protection \cite{Vlad, QEP} and similarly to the approach of dynamic decoupling \cite{DD}. Moreover, the described scheme can be extended to compensate for higher than quadratic order distortion, once more advanced tailoring of the dispersion laws is possible.

The structure of this paper is the following. In Section~\ref{SI}, we develop the method for two modes and illustrate it with a simple rectangular waveguide. The exact expressions for the mode dispersions permit us to suggest an explicit spatial distribution of ideal generic scatterers coupling the modes  and thus backing the translational invariance. Note that the term ``ideal'' abbreviates reflectionless scatterers, while the term ``generic'' has an exact meaning: its own scattering action matrix has to satisfy the so-called bracket generation condition \cite{Vlad} (implying that the canonical set of commutators holds as a complete basis for the given Hilbert-Schmidt Space). To this end, we derive $k$ non-linear algebraic equations in analytic form determining the positions $\left\{L_1,\ldots,L_M\right\}$ of $M=2k$ scatterers (making a single unit cell), such that signals on the two modes propagate in phase, synchronously and without dispersion up to Taylor expansion remainder of $k$-th order. In Section~\ref{SII}, we extend the approach to the $N$ modes dispersion problem and show that, in this case, the proper distribution of $M=N k$ generic ideal scatterers per perturbation period, can be found numerically. In Section~\ref{SIII}, we describe how the condition of ideal scatterers can be relaxed and thus reflections (back-scattering) are taken into account  by extending the theoretical techniques developed in \cite{FE}. In Section~\ref{SIV}, we implement our method in the more involved case of a $4$-mode waveguide employing non-ideal scatterers.

\section{Methodology for two modes\label{SI}}
\subsection{Formulation}
Let us consider two modes of a waveguide propagating along $z$-direction possessing the following tailored dispersion laws around an operating frequency $\omega_0$
\begin{eqnarray}
\kapa_1(\omega)\!\cong\!\kapa_1(\omega_0)\!+\!\kapa_1'(\omega_0)(\omega-\omega_0)\!+\!\frac{\kapa_1''(\omega_0)}{2}(\omega-\omega_0)^2 
\label{dispA} \\
\kapa_2(\omega)\!\cong\!\kapa_2(\omega_0)\!+\!\kapa_2'(\omega_0)(\omega-\omega_0)\!+\!\frac{\kapa_2''(\omega_0)}{2}(\omega-\omega_0)^2
\label{dispB}
\end{eqnarray}
such that our main requirement
\begin{equation}
\kapa_1''(\omega_0)=-\kapa_2''(\omega_0),
\label{req}
\end{equation}
is fulfilled. 

The propagation of such a two-mode signal at $\omega\cong\omega_0$ along a length $L$ of the waveguide, results in the accumulation of mode phase factors, which can be written in matrix form as multiplication of the two-component mode amplitude vector by a unitary matrix 
\begin{equation}
\widehat{P}_L=\exp\left[-i L\left(F_z(\omega)\widehat{\sigma}_z+F_I(\omega)\widehat{I}\right)\right],
\label{UL}
\end{equation}
where $\left\{\widehat{\sigma}_x,\widehat{\sigma}_y,\widehat{\sigma}_z\right\}$ denote the Pauli matrices \cite{QEP}, $\widehat{I}$ is the identity matrix, $F_z(\omega)=\frac{\kapa_1(\omega)-\kapa_2(\omega)}{2}$, and $F_I(\omega)=\frac{\kapa_1(\omega)+\kapa_2(\omega)}{2}$. Notably, as a result of the condition (\ref{req}) for $\omega\cong\omega_0$, the coefficient $F_I(\omega)$ does not contain the quadratic dispersion term $(\omega-\omega_0)^2$ in its Taylor expansion.

Our aim is to provide a control scheme which eliminates the effect of $F_z(\omega)$ in the vicinity of $\omega_0$; therefore, the signals encoded in two modes will be received at the same time and without dispersion. Accordingly, we introduce scatterers able to couple the modes along the line of propagation and we treat their positions as the control parameters of the problem.

Let us consider scatterers possessing the following scattering matrix
\begin{equation}
\widehat{S}=\exp\left[i F_S(\omega)\widehat{s}\right],
\label{sca}
\end{equation}
where $\widehat{s}$ is a generic traceless $2\times 2$ Hermitian matrix. This unitary model  describes only forward scattering and in Section~\ref{SIII} we show how the general case can be treated. 

In this context, our intention is to eliminate dispersion up to $2$-nd order of the Taylor expansion and thus we are using $3$ scatterers at positions $\left\{L_1,L_2,L_3\right\}$ along the propagation axis. Apparently, the total transfer matrix of this system reads:
\begin{eqnarray}
\widehat{T}=\prod\limits_{m=3}^1\left(e^{-iL_m\widehat{\sigma}_zF_z(\omega)}e^{i F_S(\omega)\widehat{s}}\right)
\end{eqnarray}
with a common phase factor $e^{-i(L_1+L_2+L_3)F_I(\omega) \widehat{I} }$. If $\widehat{T}$ equals to a nondegenerate square root of the identity matrix \cite{FE} with  accuracy of the order of $(\omega-\omega_0)^3$, then by spatially repeating twice the aforementioned triad of scatterers, the overall dispersion is canceled into the formed unit cell. Formally, this requirement implies that the linear coefficient $c_1(L_1,L_2,L_3,\omega)$ in the characteristic polynomial
\begin{equation}
{\rm Det}(\widehat{T}-\lambda\widehat{I})=\lambda^2+c_1(L_1,L_2,L_3,\omega)\lambda+1 
\label{MIII}
\end{equation}
vanishes together with its two first frequency derivatives 
\begin{eqnarray}
\left.          c_1(L_1,L_2,L_3,\omega)                    \right\vert_{\omega=\omega_{0}}  & = & 0,\nonumber\\
\left. \partial c_1(L_1,L_2,L_3,\omega)/\partial\omega     \right\vert_{\omega=\omega_{0}}  & = & 0,\nonumber\\
\left. \partial^2 c_1(L_1,L_2,L_3,\omega)/\partial\omega^2 \right\vert_{\omega=\omega_{0}}  & = & 0. 
\label{MIV}
\end{eqnarray}
In such a case, $\widehat{T}^2$ becomes the identity matrix up to a cubic correction, namely: $\widehat{T}^2=e^{\widehat{\sigma}O((\omega-\omega_0)^3)}$, where $\widehat{\sigma}$ is a $2\times2$ matrix of norm one. Therefore, both waveguide modes have identical phase shifts $2(L_1+L_2+L_3)F_{I}(\omega)$, lacking quadratic terms in their Taylor expansion over frequency; such a feature makes the signal propagate with identical group velocity in both modes and with no quadratic dispersion. Obviously, unit cells repeat themselves between transmitter and receiver however distantly they are placed. It should be also stressed that the identification of square root of unity \cite{FE} via the characteristic polynomial instead of the unity itself, permits analytic expressions.

\begin{figure}[ht!]
\centering
\subfigure[]{\includegraphics[width=5.0cm]{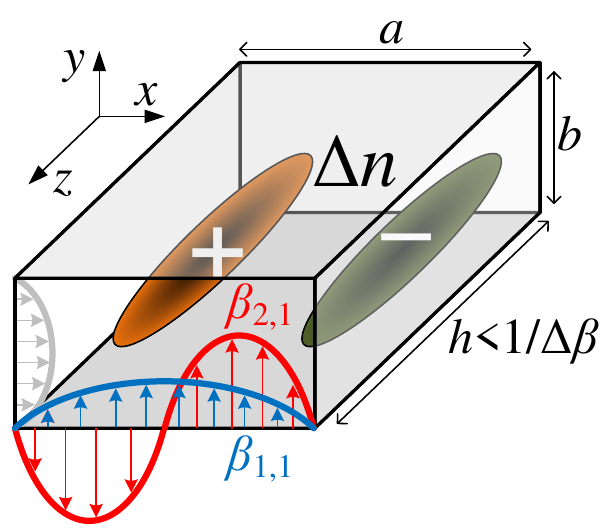}
   \label{fig:Fig1a}}\\
\subfigure[]{\includegraphics[width=3.9cm]{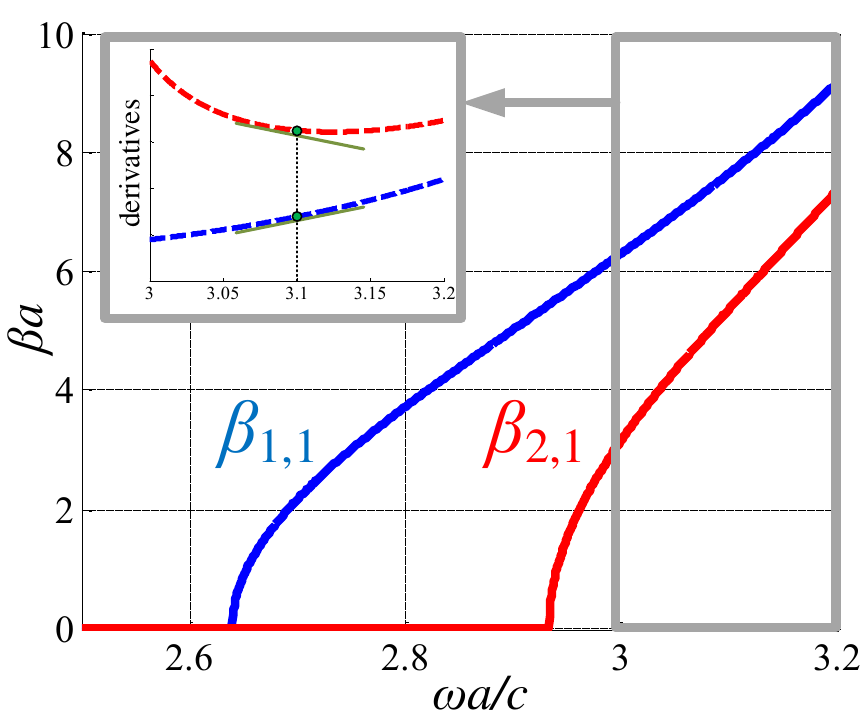}
   \label{fig:Fig1b}}
\subfigure[]{\includegraphics[width=4.3cm]{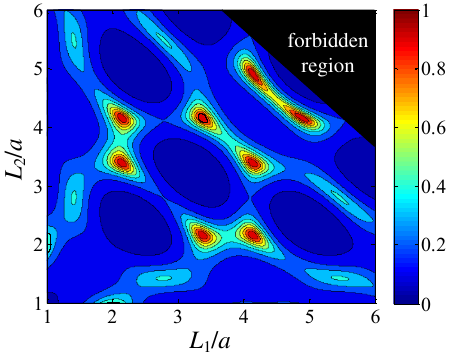}
   \label{fig:Fig1c}}
\caption{(a) Sketch of rectangular $a\times b$ waveguide with conducting walls and suitable perturbation $\Delta n$ of the refractive index. The spatial profiles of the modes with wavenumbers $\kapa_{1,1},\kapa_{2,1}$ are also depicted. (b) Dispersion law for the two first modes of rectangular waveguide ($a=2b$) for suitable inverse quadrature dispersion $n(\omega)$. The slopes of derivatives $d\kapa/d\omega$ are opposite at $\omega\cong 3.1c/a$ as shown in the inset. (c) Variation of a quantity becoming equal to one if our conditions (\ref{MIV}) are fulfilled as function of the distances $(L_1,L_2)$ normalized by waveguide width $a$.}
\label{fig:Figs1}
\end{figure}

\subsection{Example}
As an example, we consider a rectangular $a\times b$ waveguide with conducting walls, shown in Fig.~\ref{fig:Fig1a}. If we regard a frequency-dependent refraction index $n(\omega)$ for the material that fills it, then the scalar magnetic potential of the representative $(j,l)$ mode given by $A_{j,l}=\sin\left(\pi j\frac{x}{a}\right)  \sin\left(\pi l\frac{y}{b}\right)  e^{-i\kapa_{j,l}z}$, follows different dispersion laws :
\begin{equation}
\kapa_{j,l}(\omega)=\sqrt{n^2(\omega)\left(\frac{\omega}{c}\right)^2-\left(\frac{j\pi}{a}\right)^2-\left(\frac{l\pi}{b}\right)^2}.
\label{EQ1}
\end{equation}
We assume a locally inverse quadrature profile for $n(\omega)\cong 1+\frac{15}{16-(a\omega/c)^2}$ and select an operation point ($a=2b$, $\omega_0\cong3.1c/a$) where only two modes are guided: $(j=1,l=1)$ and $(j=2,l=1)$ whose spatial profiles are shown in Fig.~\ref{fig:Fig1a}. In this context, one can find a regime where the dispersion $\partial^2\kapa/\partial\omega^2$ at $\omega_0$ for the first mode is opposite to that of the second one. 

At a specific point $z=L_z$ along the propagation axis, one locates a scatterer flipping the amplitudes of the modes. This can be achieved, for instance, via antisymmetric Gaussian perturbation of the refraction index. By choosing the size $\delta n$ of the refraction index perturbation
\begin{eqnarray}
\Delta n=\delta n~e^{-\frac{(z-L_z)^2}{h^2}} e^{-\frac{y^2}{\alpha^2}}\left[e^{-x^2/\alpha^2}-e^{-(x-a)^2/\alpha^2}\right],
\end{eqnarray}
the internal products of two modes i.e., $V_{1,2} =\iiint\Delta n(x,y,z)A_{1,1}(x,y,z)A_{1,2}(x,y,z)dxdydz$ can be set in such a way that the operator $\widehat{S}$ is generated by an operator $\widehat{s}$ proportional to the Pauli matrix $\widehat{\sigma}_y$. More specifically: $\widehat{s}=F_y(\omega)\widehat{\sigma}_y$ with $F_y(\omega_0)=\pi/2$, under the assumption of smooth perturbation guaranteeing negligible back reflection. The refractive index change is schematically depicted in Fig.~\ref{fig:Fig1a} as ellipses corresponding to a scatterer being responsible for mild positive (+) and negative (-) variations $\delta n$. Such a scatterer has a length $h$ longer that the maximum mode wavelength but much shorter than the inverse of the mode wavevectors difference $\Delta\kapa=\kapa_{1,1}-\kapa_{2,1}$.

In order to find the placements of the scatterers which will eliminate dispersion, we derive the analytic expression for the linear coefficient $c_1(L_1,L_2,L_3,\omega)$ of the characteristic polynomial (\ref{MIII}),
\begin{eqnarray}
c_1(L_1,L_2,L_3,\omega)=2\left[\cos F_y(\omega)-\cos^3 F_y(\omega)\right]\nonumber\\
\times\left\{\begin{array}{c}
 \cos\left[F_z(\omega)(L_1-L_2-L_3)\right]\\
+\cos\left[F_z(\omega)(L_1+L_2-L_3)\right]\\
+\cos\left[F_z(\omega)(L_1-L_2+L_3)\right]
\end{array}\right\}\nonumber\\
-2 \cos^3F_y(\omega)\cos\left[F_z(\omega)(L_1+L_2+L_3)\right].
\end{eqnarray}
In the vicinity of $\omega\cong\omega_0$, $\cos F_y(\omega)\varpropto(\omega-\omega_0)$, and therefore, to be consistent within the $3$rd order remainder approximation, the term proportional to $\cos^3 F_y(\omega)$ has to be ignored. In this way, $c_1$ vanishes automatically at $\omega=\omega_0$ and the demand for zero derivatives (\ref{MIV}) give two equations with three unknowns $\left\{L_1,L_2,L_3\right\}$. Accordingly, a continuous family of solutions can be determined numerically, which is trivially discretized after imposing the additional physical requirement $(L_1+L_2+L_3)F_I(\omega_0)=2v\pi$ for integer $v$. Apparently, the obtained lengths should be positive quantities, otherwise one more free parameter (additional scatterer) may be added. For a specific setup, namely $v=10$, $\omega_0\cong3.1c/a$, we obtain $F_z(\omega_0)\cong 1.14/a$, $F_I(\omega_0)\cong 6.50/a$ and we numerically find one of many possible solutions for the locations of the three scatterers ($L_1+L_2+L_3\cong 9.67a$$\Rightarrow$$L_3\cong 9.67a-L_1-L_2$). For this example, one can define the quantity $\frac{\e}{|c_1'(\omega_0)|^2+|c_1''(\omega_0)|^2+\e}$ for small $\e>0$, which reaches unitary value only if all the conditions (\ref{MIV}) are fulfilled.  Note, we obtain $c_1(\omega_0)=0$ automatically since $F_y(\omega_0)=\pi/2$. In Fig. \ref{fig:Fig1c}, we represent the aforementioned quantity on the map $(L_1/a,L_2/a)$, while the combinations giving $L_3<0$ are labeled as ``forbidden region''. It is easily observed that there are multiple (eight of them are depicted in the considered parametric box) pairs of distances $(L_1,L_2)$ giving designs that satisfy the constraints (\ref{MIV}). If one chooses $L_1\cong 3.38a$ and $L_2\cong 4.15a$, one obtains the following transfer matrix for a single unit cell:
\begin{eqnarray}
\widehat{T}^2\cong 
e^{i 10^{-5}\!(3.4\widehat{\sigma}_x\!+\!4.7 \widehat{\sigma}_y\!+\!0.17 \widehat{\sigma}_z)(\omega-\omega_0)^3(a/c)^3},
\label{EQ14}
\end{eqnarray}
with a common phase factor proportional to $e^{i \left[166(\omega-\omega_0)a/c+420(\omega-\omega_0)^3(a/c)^3\right]\widehat{I}}$. Indeed, almost dispersion-free propagation (up to $3$-rd Taylor order) is achieved.

\section{Generalizing to $k$ Taylor order and to $N$ modes\label{SII}}
To better ensure the identity of propagating signals in different modes, the order $k$ of the Taylor remainder can be increased at will. In particular, the system (\ref{MIV}) for such a case has to be extended to incorporate more scatterers and requires zero higher-order derivatives up to $\partial^{k-1}c_1/\partial\omega^{k-1}$ at $\omega=\omega_0$. Eliminating the higher order terms also in the common phase of the modes, requires additional tailoring of the frequency-dependent $n(\omega)$, potentially with help from the paradigm of metamaterials \cite{MetaMat}.

We now  consider the general formulation of the problem in the case of $N$ modes with different dispersion laws $\kapa_{l}(\omega)$. The initial aim is to achieve a breaking of translational symmetry so that all modes have identical dispersions up to the $k$-th order correction of their Taylor expansion. As earlier, we consider a sequence of scattering transformations
\begin{equation}
\widehat{T}\left(\left\{L_m\right\}  ,\omega\right)  =
\prod \limits_{m=M}^1\left(e^{-iL_m\widehat{\phi}(\omega)}e^{i\widehat{\chi}(\omega)}\right)
\label{EQ15}
\end{equation}
where $\widehat{\chi}(\omega)$ is a generic frequency dependent $N\times N$ Hermitian scattering action matrix, and $\widehat{\phi}(\omega)$ is a real traceless diagonal $N\times N$ matrix. The elements of $\widehat{\phi}(\omega)$ are given by $\phi_{ll}(\omega)=\kapa_{l}(\omega)-\frac{1}{N}\sum_{j=1}^{N}\kapa_j(\omega)$ describing the mode phase difference accumulation per unit waveguide distance. Note that the essential number of scatterers $M$ is suitably dependent both on the number of modes $N$ and the accuracy order $k$. The average mode wavevector $\kapa_0(\omega)=\frac{1}{N}\sum_{j=1}^{N}\kapa_j(\omega)$ contributes with a global phase $\kapa_0(\omega)\sum_{m=1}^M L_m$, which is factored out from the product (\ref{EQ15}). By analogy to (\ref{req}), we can assume no second order dispersion, namely $\kapa_0''(\omega_0)=\frac{1}{N}\sum_{j=1}^N\kapa_j''(\omega_0)=0$ and try to identify the length intervals $\left\{L_m\right\}$ such that $\widehat{T}$ is a nondegenerate $N$-th root of the identity operator $\widehat{I}$ up to $O((\omega-\omega_0)^k)$ term.

\begin{figure}[ht!]
\centering
\subfigure[]{\includegraphics[width=8.0cm]{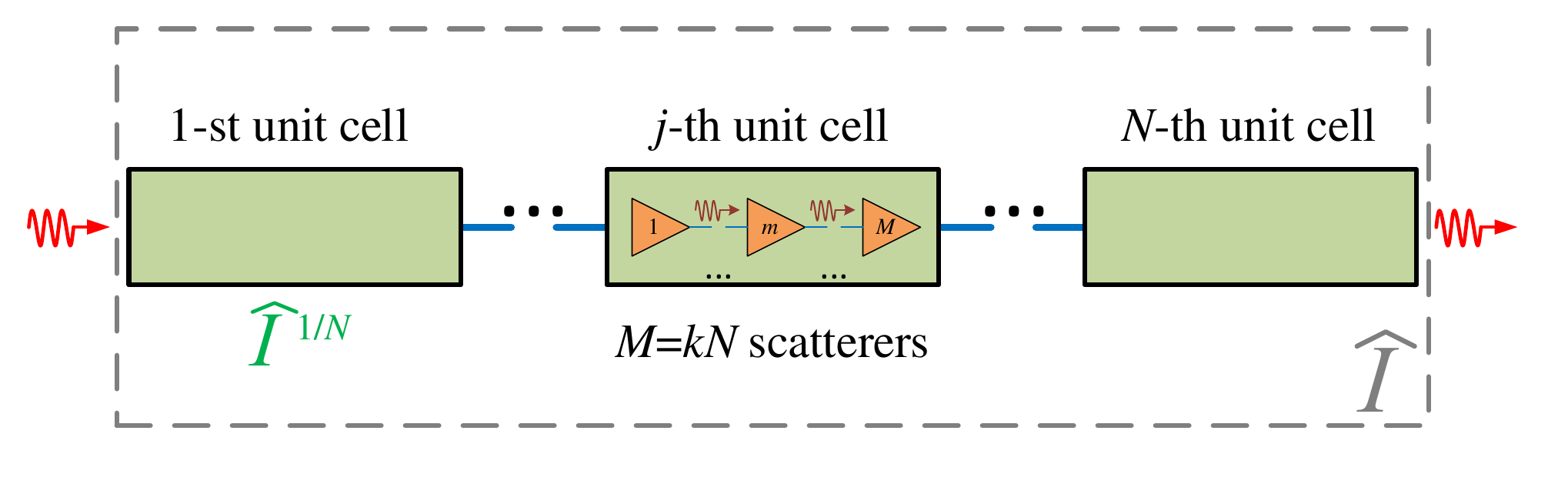}
   \label{fig:NewFigA}}
\subfigure[]{\includegraphics[width=8.0cm]{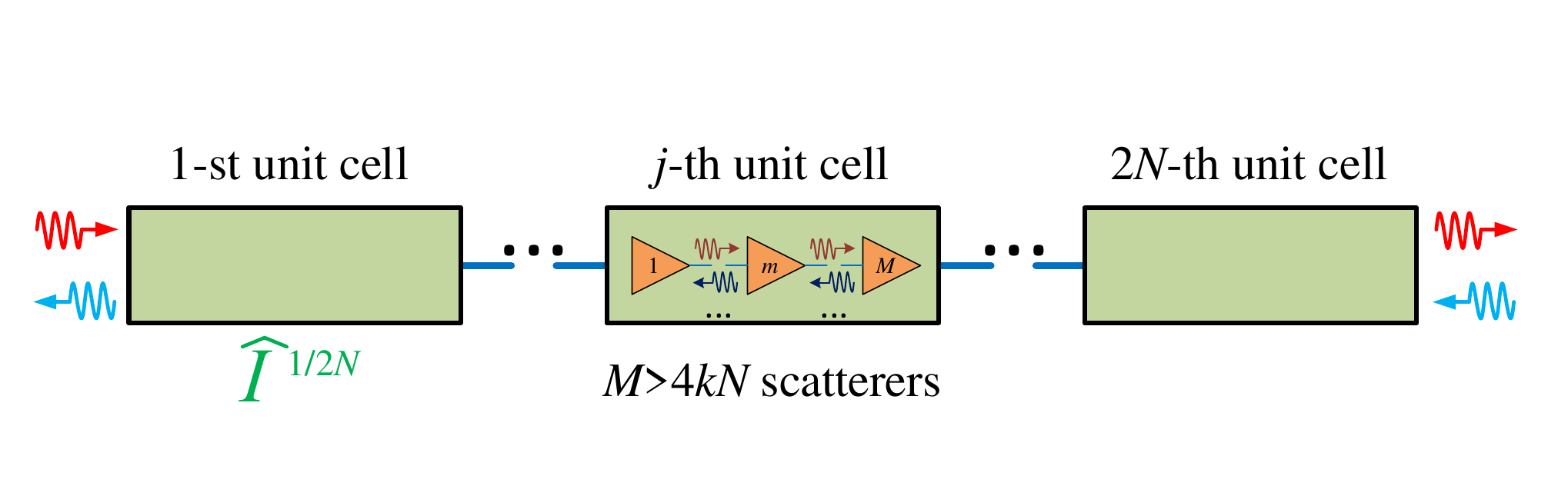}
   \label{fig:NewFigB}}
\caption{ Depictions of the proposed dispersionless propagation link comprised by multiple unit cells, each of which contains several scatterers.  
 (a) In the absence of reflections $N$ unit cells are  required and $M=kN$ scatterers per unit cell. Each
unit cell has a scattering matrix equal to $\widehat{I}^{1/N}$, making an overall unitary response $\widehat{I}$.  (b) In the presence of reflections (back-scattering) requiring $2N$ unit cells (double number of modes due to the oppositely propagated waves) and at least $M\ge 4kN$ scatterers per unit cell.}
\label{fig:FigsAB}
\end{figure}

To this end, we consider the characteristic polynomial
\begin{equation}
{\rm Det}(\widehat{T}-\lambda\widehat{I})  =\lambda^{N}+\sum_{j=1}^{N-1}c_j\left(\left\{L_m\right\},\omega\right)
\lambda^j+e^{i\varphi\left(\left\{L_m\right\}  ,\omega\right)}
\label{EQ16}
\end{equation}
and require, that at $\omega=\omega_0$ all coefficients $\{c_j\}$ vanish together with all their (numerically determined) derivatives up to $\left(k-1\right)  $-th order. In addition, we demand that $\varphi\left(\left\{L_m\right\},\omega_0\right)=2v\pi$, where $v$ is an integer number (with $v<N$ to avoid degenerate roots of unity \ref{FE}), and again zero derivatives for its first $(k-1)$ Taylor orders. All together we impose $M=kN$ conditions on $kN$ variables $\left\{L_m\right\}$ and obtain an $M\times M$ system of nonlinear algebraic equations to be solved numerically for given $\widehat{\phi}(\omega)$ and $\widehat{\chi}(\omega)$. Once a solution is found, the repetition of the scattering sequence $N$-times, corresponding to the matrix $\widehat{T}^{N}(\{L_m\},\omega)=e^{O((\omega-\omega_0)^k)}$ results in the required propagation with the aforementioned common phase $N\kapa_0(\omega)\sum_{m=1}^M L_m$.  In Fig.~\ref{fig:NewFigA} we provide an abstract depiction of the proposed dispersionless concept.

\section{Incorporating  back scattering \label{SIII}}
Up to now, the method assumes ideal scatterers for mode coupling and control. However, it requires a substantial amount of scatterers along the transmission line; thus, even if the reflections are negligible for a single scatterer, the cumulative effect cannot be ignored. For this reason, we consider here the general case where the scattering matrix $\widehat{S}$ connecting the two incoming $N$-mode signals with the corresponding outgoing ones obeying only to the requirement $|{\rm Det}\widehat{S}|=1$.  Accordingly the transfer matrix $\widehat{T}_S$, corresponding to $\widehat{S}$, only satisfies $|{\rm Det}\widehat{T}_S|=1$ without necessarily being a unitary matrix. In this scenario, we need at least four times ($4kN$) more scatterers compared to the reflectionless case ($kN$) in order to achieve a similar-quality outcome. That is because we have not only forward but also backward modes developed and additionally not only the phase but also the magnitude of the transmission coefficients should also be engineered.

If $M\ge 4kN$ scatterers are placed at different positions, separated by distances/lengths $L_m$, the total (non-unitary) transfer matrix $\widehat{T}$ of the $\frac{1}{2N}$ part of the unit cell is given by:
\begin{equation}
\widehat{T}=\prod\limits_{m=M}^1\left(\widehat{T}_m\widehat{T}_S\right)
\label{UnitCellT}.
\end{equation}
Our aim is to identify such lengths $\left\{L_m\right\}$ so that the  total transmission matrix $\widehat{T}$ is equal, up to some approximation, to $\widehat{T}\cong \widehat{I}^{\frac{1}{2N}}$. Therefore, the overall transfer matrix $\widehat{T}^{2N}$ of the unit cell will be proportional to identity with a common phase factor.

The positions $\left\{L_m\right\}$ of the scatterers are determined in a similar way as in the reflection-less system (introduced in Section \ref{SII}). However, when back-scattering is not negligible, we have to simulate an overall unitary transfer matrix (identity matrix with approximation $(\omega-\omega_0)^k$) by using non-unitary blocks. It should be also stressed that the quantum control analysis \cite{FE}, invoked in this current work, does not hold for non-unitary operations and there is no guarantee on a ``perfect'' outcome of the method. For this reason, we introduce a performance indicator of the derived solution showing how close to unitary is the obtained transfer matrix. In particular, we define a quantity indicating the non-unitary character of a matrix $\widehat{X}$: $g(\widehat{X})=\sum_{l}|\lambda_l(\widehat{I}-\widehat{X}\widehat{X}^{\dagger})|$, where ${\lambda_l}(\widehat{D})$ is the $l$-th eigenvalue of the matrix $\widehat{D}$; when $\widehat{D}$ is unitary, $g(\widehat{D})=0$. In this way, one can evaluate it for the total matrix $g(\widehat{T}^{2N})$ and compare it with $g(\widehat{T}_S)$ referring to the transfer matrix of the single scatterer, which is the cause of reflections. If the former quantity is much smaller (say, one order of magnitude) than the latter one, we may claim that the method yields an acceptable result, since the whole system (comprising multiple elements) is much less non-unitary than a single element. Similarly to Fig.~\ref{fig:NewFigA}, in Fig.~\ref{fig:NewFigB} we depict the presented propagation link in the presence of reflections and thus two times more modes are appeared requiring $2N$ unit cells for an overall dispersion-free transmission $\widehat{I}$; as explained above, each unit cell should incoroporated $M\ge 4kN$ imperfect scatterers for a good approximation of the necessary scattering matrix $\widehat{I}^{1/2N}$.

\begin{figure}[ht!]
\centering
\subfigure[]{\includegraphics[width=4.1cm]{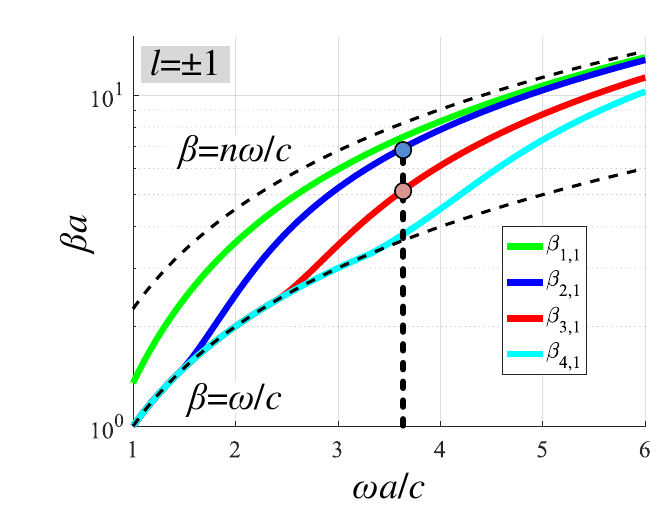}
   \label{fig:Fig2a}}
\subfigure[]{\includegraphics[width=4.1cm]{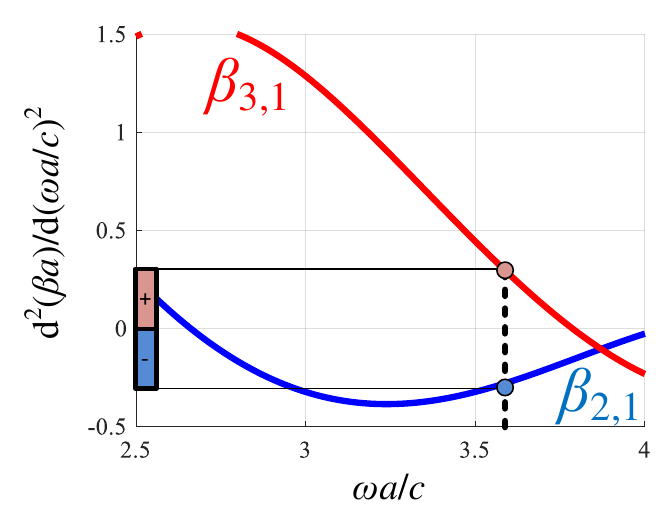}
   \label{fig:Fig2b}}
\subfigure[]{\includegraphics[width=4.1cm]{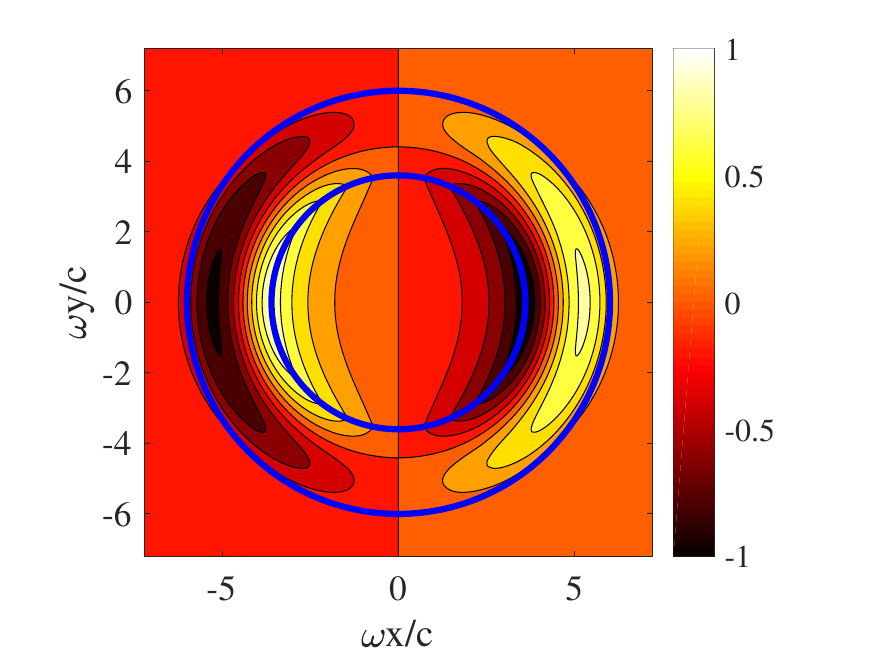}
   \label{fig:Fig2c}}
\subfigure[]{\includegraphics[width=4.1cm]{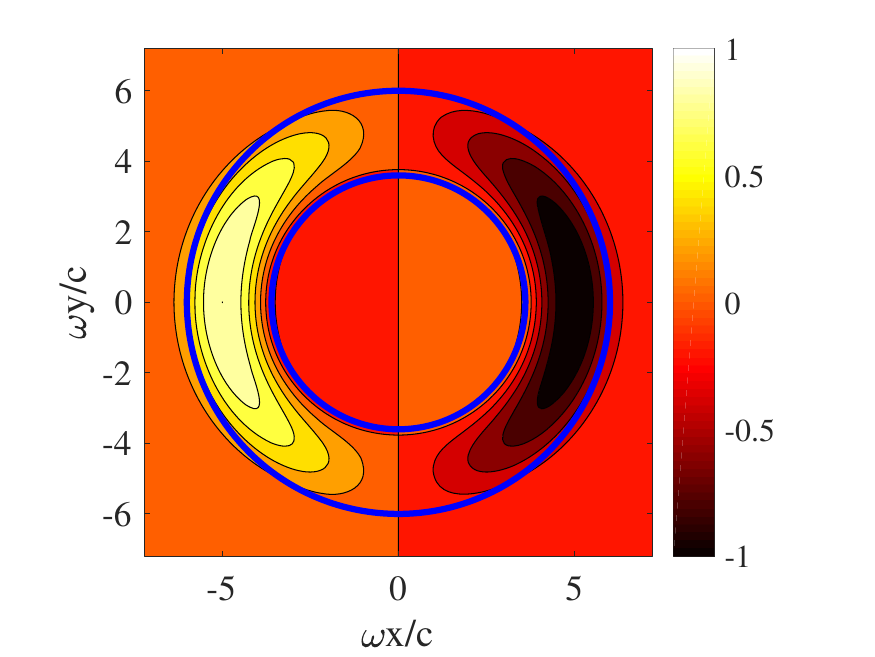}
   \label{fig:Fig2d}}
\caption{(a) The dimensionless wavenumber $\kapa a$ as function of $\omega a/c$ corresponding to the first four supported modes by an unperturbed hollow tube waveguide with internal radius $a$, external radius $b$ and refractive index $n$. (b) The dispersion $\frac{d^2 \kapa/d\omega^2}{a/c^2}$ of two modes giving opposite values at the frequency $\omega_0$ indicated by vertical dashed line. (c, d) Spatial profiles of the $z$ electric components of the employed modes at constant-$z$ plane: $\beta_{2,1}$ (Fig. \ref{fig:Fig2c}) and $\beta_{3,1}$ (Fig. \ref{fig:Fig2d}).}
\label{fig:Figs2}
\end{figure}

\section{Non-ideal scattering based on tube waveguide perturbed by twists \label{SIV}}

\subsection{Model analysis}
Let us consider a realization of the idea in a waveguide that is more involved compared to the rectangular two-mode metallic pipe, but still mostly analytically tractable. We assume a hollow circular dielectric cylinder (of refractive index $n$) with inner radius $a$ and outer radius $b$ whose supported modes ($e^{+i\omega t-i\kapa z}$) can be easily determined. In Fig.\ref{fig:Fig2a}, we consider $n\cong 2.26$ (Polyethylene at $\omega\cong(2\pi)3$ GHz), $b=5a/3$ and show the dependence of $\kapa a$ as function of $\omega a/c$ for the first four modes of the device; all the curves lie within the two dashed lines $\kapa=\omega/c$ and $\kapa=n\omega/c$. In particular, we consider first order ($l=\pm 1$) azimuthal $e^{\pm il\theta}$ profile, while the first subscript ($j$) in the wavenumber $\kapa_{j,l}$ denotes the serial number of the mode. In Fig. \ref{fig:Fig2b}, we represent the normalized dispersion $\frac{c^2}{a}\frac{d^2\kapa}{d\omega^2}$ with respect to $\omega a/c$ for two modes ($\kapa_{2,1}$, $\kapa_{3,1}$) and we clearly see that for specific frequencies $\omega$, one can achieve opposite second derivatives for propagation wavenumbers $\kapa$. For example at $\omega_0\cong 3.6c/a$ (indicated by black vertical dashed line  in both Fig. \ref{fig:Fig2a} and in Fig. \ref{fig:Fig2b}), we find for the second ($\kapa_{2,1}\cong 6.83/a$) and the third mode ($\kapa_{3,1}\cong 5.02/a$) that the required condition is fulfilled: $d^2\kapa_{3,1}/d\omega^2\cong -d^2\kapa_{2,1}/d\omega^2 \cong 0.28a/c^2$. In Figs \ref{fig:Fig2c} and \ref{fig:Fig2d}, we show the normalized transversal profiles of the (axial) electric field of the used modes at a constant-$z$ plane. Since at the considered frequency, the dispersions are opposite, alternating power along z axis between these two modes will suppress the overall signal distortion.

In order to implement the proposed dispersion control technique in the a real-world analog of the aforementioned configuration, we choose the role of the scatterers to be played by mild deformations in the shape of the waveguide's cylindrical boundaries. Such a structural imperfection mixes the guided modes towards both directions (forward and backward) and provides low level of reflections. In the following, we are not rigorously solving  the formulated boundary value problem but only heuristically obtain the effective scattering matrices incorporated in the proposed method. In particular, we show the key algebraic steps for deriving an approximate solution providing us with the scattering characteristics of these mild ridges, while capturing the basic physics of the problem. For this reason, we avoid including all the cumbersome expressions and referring to all the relevant details or assumptions leading to the approximate solution, since it goes beyond the scope of this study. Therefore, one can employ a perturbed cylindrical coordinate system (with small parameter $\gamma$) at which the Laplace operator is no longer just a sum of second derivatives but includes more terms proportional to strength $\gamma$. Into such a system, the Cartesian $(x,y)$ coordinates are written in the form:
\begin{eqnarray}
x  &  =[r+\gamma p(r,z)\cos(2\theta+\psi(z)z)] \nonumber\\
   & \times \cos\left(\theta-\gamma q(r,z)\sin(2\theta+\psi(z)z)\right) \nonumber\\
y  &  =[r+\gamma p(r,z)\cos(2\theta+\psi(z)z)] \nonumber\\
   & \times \sin\left(\theta-\gamma q(r,z)\sin(2\theta+\psi(z)z)\right),
\end{eqnarray}
where $(r,\theta,z)$ are the cylindrical coordinates and $\psi(z)$ is a pseudo-wavevector function stating how the initial  phase is changed along the $z$ axis. The functions $p,q$ are then chosen in such a way, that the Jacobian remains equal to $r$ (as in the simple cylindrical case) up to the second order terms in $\gamma$. Such a condition is equivalent to
\begin{eqnarray}
\frac{\partial p(r,z)}{\partial r}+\frac{p(r,z)}{r}-2q(r,z)=0.
\end{eqnarray}

One can assume separable form for functions $p,q$, namely $p(r,z)=Z(z)R(r)$ and accordingly $q(r,z)=\frac{Z(z)}{2}\left(R'(r)+\frac{R(r)}{r}\right)$. We can choose as $R(r)$ a function creating a symmetric perturbation of the radial boundaries proportional to $\gamma$ around their average values $r=a,b$, namely $R(r)=\left(r-\frac{a^2+b^2}{a+b}\right)r$. Furthermore, we take a linear phase change along $z$ axis with a constant pseudo-wavevector $\psi(z)=\psi$, but most importantly we adopt an oscillating profile for $Z(z)$ possessing an envelope vanishing far from $z=0$, which is the scatterer's position. More specifically, we pick $Z(z)=Q\frac{\sin(uz)}{\sinh(\frac{\pi z}{2w})}$, whose Fourier transform ($z  \rightleftharpoons  k_z$) has the simple form: $\mathcal{F}\left\{Z(z)\right\}\sim Qw \left[\tanh((k_z+u)w)-\tanh((k_z-u)w)\right]$. 

\begin{figure}[ht!]
\centering
\subfigure[]{\includegraphics[width=3.8cm]{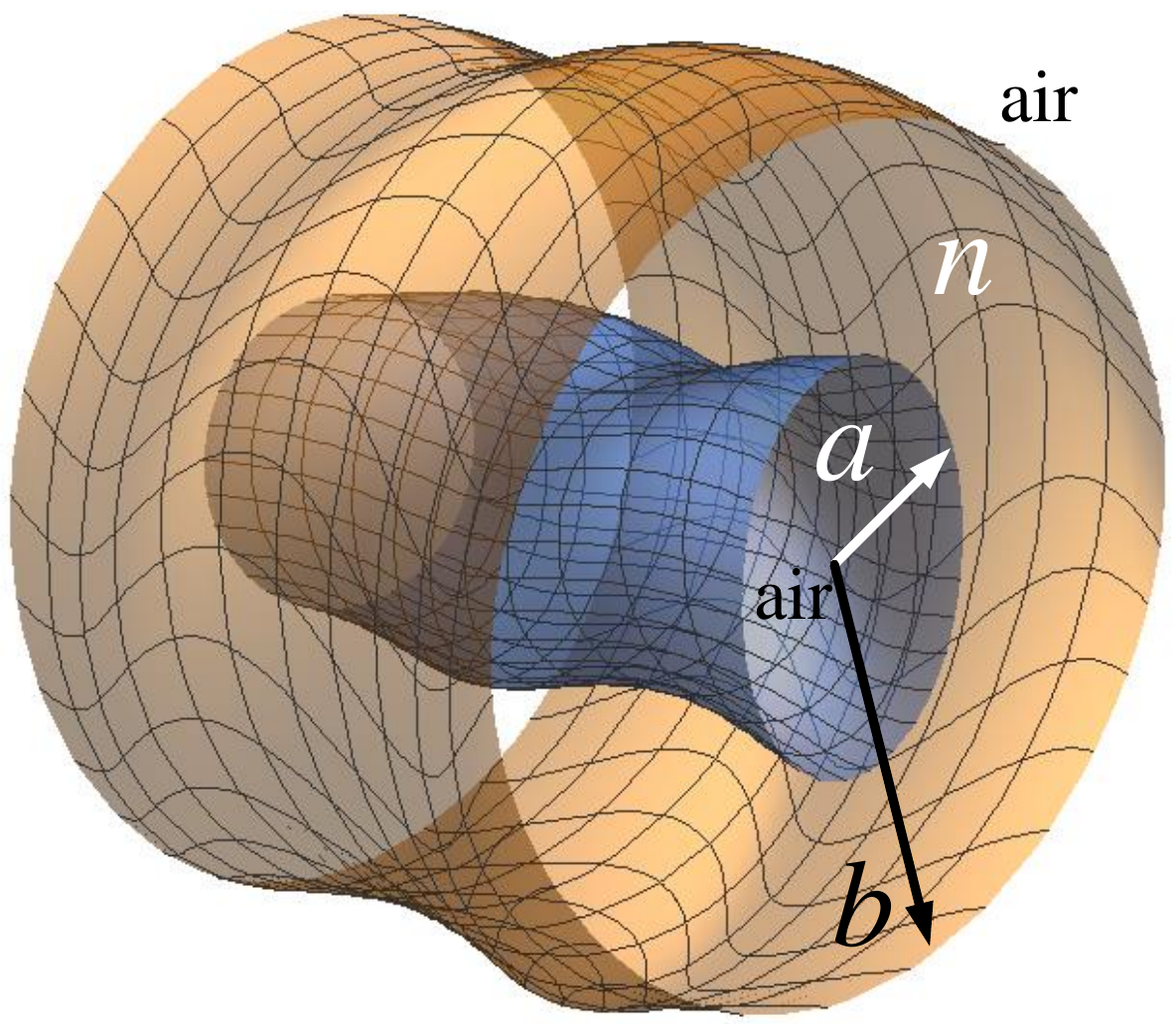}
   \label{fig:Fig3a}}
\subfigure[]{\includegraphics[width=4.4cm]{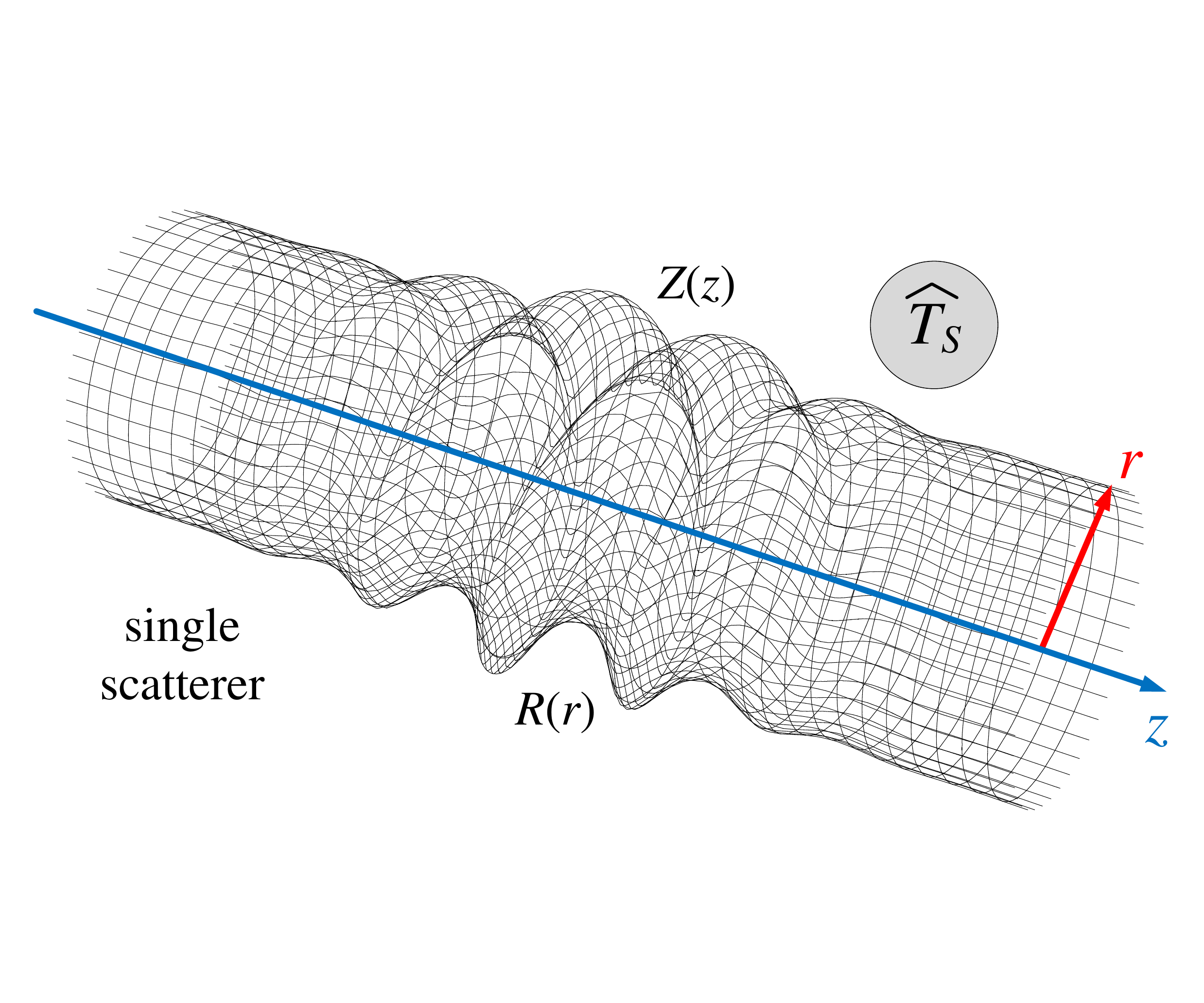}
   \label{fig:Fig3b}}
\caption{Sketches of: (a) A small segment of the perturbed waveguide with average internal radius $a$ and external $b$, which is hollow and filled with material of refractive index $n$. (b) The coordinate system, for a fixed radius $r$, describing a single scatterer (with transfer matrix $\widehat{T}_S$) whose net perturbation at both its surfaces (internal or external) disappears far from it.}
\label{fig:Figs3}
\end{figure}

In Fig. \ref{fig:Fig3a}, we show a segment of the perturbed waveguide with average boundaries $r=a,b$ filled with dielectric material of non-dispersive refractive index $n$ for the aforementioned selection of functions $\left\{R(r), Z(z), \psi(z)\right\}$. The perturbed coordinate lines across a longer part of the waveguide corresponding to a single scatterer (with matrix $\widehat{T}_S$) that provides mode coupling with negligible reflections, a depicted in Fig. \ref{fig:Fig3b}.

In order to approximately evaluate the transfer matrix of the scatterer we use the modes of the well-established unperturbed cylindrical waveguide (of Fig. \ref{fig:Fig2a}) with $z$-dependent, slowly varying amplitudes and plug them into the Helmholtz equation. To this end, we project the obtained expressions to the same set of modal profiles and integrate over the cross section of our tube \cite{Marcuse, Pert}. In this way, the mode of opposite angular momentum order $\pm l$ are getting coupled unlike what is happening in the unperturbed analog. That is because the geometry of the waveguide (as shown in Fig. \ref{fig:Fig3b}) discerns the right-handed $e^{-il\theta}$ propagation from the left-handed $e^{+il\theta}$ propagation.

\subsection{Transfer matrix evaluation}
\label{Teval}
Let us work with the two modes $\kapa_{2,1}$ and $\kapa_{3,1}$ of Fig. \ref{fig:Fig2a} and their opposite angular momentum counterparts ( $\kapa_{2,-1}$ and $\kapa_{3,-1}$) exhibiting opposite dispersion at $\omega\cong\omega_0$ as demonstrated at Fig. \ref{fig:Fig2b}; note that the total number of modes is eight because both directions of propagation are taken into account. The $8\times 8$ transfer matrix of the scatterer of Fig. \ref{fig:Fig3b} has the form:
\begin{equation}
\widehat{T}_S=\exp\left\{i\left[\begin{array}{cccc}
\widehat{O}             & \widehat{X}^{FF}_{1,-1} & \widehat{O}             & \widehat{X}^{FB}_{1,-1} \\
\widehat{X}^{FF}_{-1,1} & \widehat{O}             & \widehat{X}^{FB}_{-1,1} & \widehat{O} \\
\widehat{O}             & \widehat{X}^{BF}_{1,-1} & \widehat{O}             & \widehat{X}^{BB}_{1,-1} \\
\widehat{X}^{BF}_{-1,1} & \widehat{O}             & \widehat{X}^{BB}_{-1,1} & \widehat{O}
\end{array}\right]\right\},
\label{TransferMatrixForm}
\end{equation}
where $\widehat{O}$ is the $2\times 2$ zero matrix and the rest of the $2\times 2$ blocks $\widehat{X}$ have superscripts indicating forward ($F$) or backward ($B$) waves and subscripts referring to right-handedness ($l=-1$) or left-handedness ($l=+1$) of the corresponding modes. 

We consider the same parameter set defining the unperturbed configuration used in Figs \ref{fig:Figs3} namely: $n\cong 2.26$, $b=5a/3$, $\omega_0\cong 3.6c/a\cong (2\pi)3$ GHz and suitable values for the perturbed waveguide boundaries, namely: $u=2$, $w=4$, $\psi=1.78$, $Q=0.2$. In such a scenario, the $2\times 2$ blocks $\widehat{X}$ of (\ref{TransferMatrixForm}) have explicit forms given by (\ref{eq:FMatrix}) and (\ref{eq:BMatrix}) with the numerical values directly stemming from the coupling analysis discussed above. Note that the transfer matrix $\widehat{T}_S=\widehat{T}_S(\omega)$ is frequency-dependent since $\W=\frac{\omega-\omega_0}{\omega_0}$ is the relative frequency shift, while the auxiliary functions of $\W$ used in (\ref{eq:FMatrix}), (\ref{eq:BMatrix}) are defined as follows: $\Phi_{\W}=\W(\W-61.65)-55.95$, $\Gamma_{\W}=\W(\W+68.53)+45.46$, $F_{\W}(x,y)=\cosh(\W(0.22\W+x)+y)+3.76$ and $G_{\W}(x)=\cosh(14.70\W+x)+3.76$.

\begin{widetext}
\begin{eqnarray}
\left[
\begin{array}{cc}
\widehat{X}^{FF}_{1,-1} & \widehat{X}^{FB}_{1,-1} \\
\widehat{X}^{FF}_{-1,1}	& \widehat{X}^{FB}_{-1,1} 
\end{array}
\right]\cong\left[
\begin{array}{cccc}
 \frac{-0.73i}{\Ph}                         & \frac{-1.94-20.95i}{F_{\W}(0.77,2.33)\Ph}  &  \frac{-15.15i}{F_{\W}(-13.92,-9.11)\Ph}  & \frac{-1.94-20.95i}{G_{\W}(7.93)\Ph}   \\
 \frac{-1.94+20.95i}{F_{\W}(0.77,-4.70)\Ga} & \frac{1.22i}{\Ga}                          &  \frac{-1.94+20.95i}{G_{\W}(7.93)\Ga}     & \frac{25.34i}{F_{\W}(15.47,6.74)\Ga}   \\
 \frac{-2.75i}{\Ph}                         & \frac{-1.94-20.95i}{F_{\W}(0.77,-0.03)\Ph} &  \frac{-15.15i}{F_{\W}(-13.92,-11.48)\Ph} & \frac{-1.94-20.95i}{G_{\W}(10.30)\Ph}  \\
 \frac{-1.94+20.95i}{F_{\W}(0.77,-2.33)\Ga} & \frac{4.60i}{\Ga}                          &  \frac{-1.94+20.95i}{G_{\W}(10.30)\Ga}    & \frac{25.34i}{F_{\W}(15.47,9.11)\Ga}   
\end{array}
\right],
\label{eq:FMatrix}
\end{eqnarray}
\begin{eqnarray}
\left[
\begin{array}{cc}
\widehat{X}^{BF}_{1,-1} & \widehat{X}^{BB}_{1,-1} \\
\widehat{X}^{BF}_{-1,1}	& \widehat{X}^{BB}_{-1,1} 
\end{array}
\right]\cong\left[
\begin{array}{cccc}
\frac{15.15i}{F_{\W}(-13.92,-16.15)\Ph} & \frac{1.94+20.95i}{G_{\W}(14.96)\Ph}   & \frac{0.73i}{\Ph}                         & \frac{1.94+20.95i}{F_{\W}(0.77,-4.70)\Ph} \\
\frac{1.94-20.95i}{G_{\W}(14.96)\Ga}    & \frac{-25.34i}{F_{\W}(15.47,13.78)\Ga} & \frac{1.94-20.95i}{F_{\W}(0.77,2.33)\Ga}  & \frac{-1.22i}{\Ga}                        \\
\frac{15.15i}{F_{\W}(-13.92,-13.78)\Ph} & \frac{1.94+20.95i}{G_{\W}(12.60)\Ph}   & \frac{2.75i}{\Ph}                         & \frac{1.94+20.95i}{F_{\W}(0.77,-2.33)\Ph} \\
\frac{1.94-20.95i}{G_{\W}(12.60)\Ga}    & \frac{-25.34i}{F_{\W}(15.47,11.41)\Ga} & \frac{1.94-20.95i}{F_{\W}(0.77,-0.03)\Ga} & \frac{-4.60i}{\Ga}  
\end{array}
\right]
\label{eq:BMatrix}.
\end{eqnarray}
\end{widetext}

Note that the model of Figs \ref{fig:Fig3a}, \ref{fig:Fig3b} represented by the perturbed matrix of the single cell \eqref{TransferMatrixForm} comprising \eqref{eq:FMatrix}, \eqref{eq:BMatrix} does not couple modes of same angular momentum order $l$. Such a property constitutes a major snag in implementing the proposed method since it is based on the concept of blending the supported modes in all non-degenerate ways. Ideally, such a weak point may be remedied by placing before the scatterer an additional object, like a segment with longitudinally-changing cross section or a gyrotropic slab, that realizes the missing coupling. However, since we are not analyzing the corresponding setup of Figs \ref{fig:Figs3} rigorously anyway, we emulate the effect of the aforementioned objects by inserting small non-zero elements at the positions we need them in the matrix. An additional deficiency of the obtained transfer matrix is its tiny mixing angles (between different modes), which require an extremely large number of scatterers for a successful implementation of the proposed method. Therefore, in order to present a numerical example with a tractable number of scatterers per unit cell, we artificially boost the unitary character of the derived transfer matrix. 

\subsection{Method implementation}
\label{Mimplem}
After evaluating the matrix of the single scatterer $\widehat{T}_S(\W)$ according to \eqref{TransferMatrixForm} and performing the aforementioned modifications, we can find the total transfer matrix $\widehat{T}=\widehat{T}(\W)$ according to (\ref{UnitCellT}), which is additionally dependent on the lengths $\{L_m\}$ for $m=1,\ldots,M$. Since we consider dispersion-free approximation up to $k=3$ Taylor order correction and employing $N=4$ modes (for each direction), we need at least $4kN=48$ scatterers per unit cell. For reasons related to  satisfyingly imposing the constraints around $\w=w_0$, our choice is $M=54$. To this end, we regard the characteristic polynomial ${\rm Det}(\widehat{T}-\lambda\widehat{I})=\lambda^8 + \sum_{j=0}^7c_j \lambda^j$, where $|c_0|=1$; note that the coefficients $c_j=c_j(\{L_m\}, \W)$ for $j=1,\cdots,7$, are also functions of the lengths between the scatterers $\{L_m\}$ for $m=1,\ldots,M$ and the relative frequency shift $\W$. In this way, we obtain the objective functional ${\bf Obj}$ on $\{L_m\}$ parameters
\begin{equation}
{\bf Obj}=\sum_{j=1}^3\left\{\begin{array}{c}W|c_j(\{L_m\},0)|+\frac{\partial |c_j(\{L_m\},0)|}{\partial \W}\\
+\frac{\partial^2 |c_j(\{L_m\},0)|}{\partial \W^2}\end{array}\right\},
\label{ObjectiveFunctional}
\end{equation}
where $W$ is a weight factor of two decimal orders.

\begin{figure}[ht!]
\centering
\subfigure[]{\includegraphics[width=4.1cm]{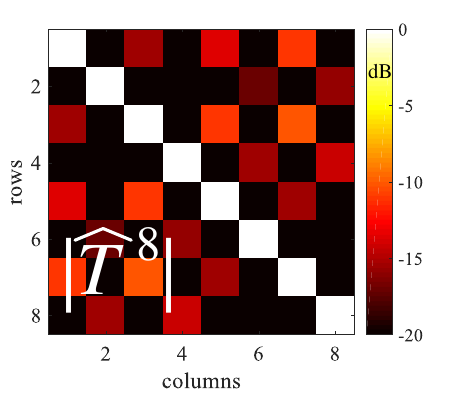}
   \label{fig:Fig4a}}
\subfigure[]{\includegraphics[width=4.1cm]{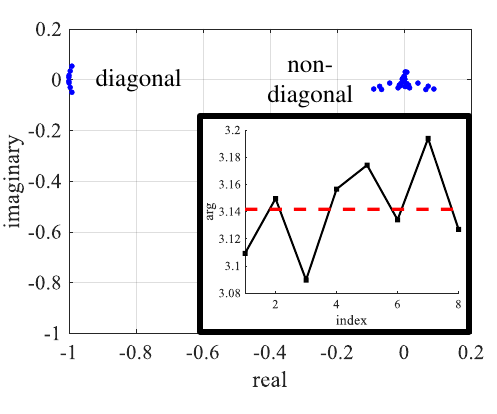}
   \label{fig:Fig4b}}
\caption{The optimal solution of scatterers positions minimizing (\ref{ObjectiveFunctional}) for the numerical setup of Subsections \ref{Teval} and \ref{Mimplem} applied to the model of Figs \ref{fig:Figs3} at $\omega=\omega_0 \Rightarrow \Omega=0$. (a) The magnitudes (in dB on column/row map) of the elements of the (eighth power of) unit cell total transfer matrix $\widehat{T}^8$ at the central operating frequency $\omega=\omega_0$. (b) Representation of the elements of the same matrix on the complex plane, where two clusters are formulated: one corresponding to the diagonal and another to the non-diagonal elements. Therefore, the unit cell very mildly distorts  the signal, while the reflections are almost negligible. The variation of the phase of the diagonal elements is shown in the inset where the $\pi$ value is remarked by a dashed line.}
\label{fig:Figs4}
\end{figure}

We perform a random search ($\sim 10^5$ trials) on different sets of lengths $\{L_m\}$ and we select the ones for which the metric ${\bf Obj}$ of (\ref{ObjectiveFunctional}) falls below a specific threshold. By using these sets as initial points, we continue with a gradient method on the objective functional in the space of length parameters, where the derivatives are numerically calculated at each step. Eventually, among all final points of all trajectories in the $\{L_m\}$ space, we select the point reaching the lowest final value for ${\bf Obj}(\{L_m\})$. Therefore, the optimal solution for the lengths is obtained and the transfer matrix  $\widehat{T}$ of the unit cell is close to identity $\widehat{I}$ (with a common phase $\pi$). Such an outcome is demonstrated in Fig. \ref{fig:Fig4a}, where the magnitude of the elements of the matrix $\widehat{T}^N=\widehat{T}^8$ is shown in dB (logarithmic scale, $10\log_{10}|\widehat{T}^8|$) as function of the respected row and column. Indeed, the elements across the diagonal are of unitary magnitude whereas the values fall rapidly away from it. In Fig. \ref{fig:Fig4b}, we  represent the elements of $\widehat{T}^8$ on the complex plane, where two clusters of dots are formulated. The first set of points is concentrated around the value $(-1)=e^{i\pi}$ and correspond to the diagonal elements of the matrix; on the contrary, all the other elements are crammed around the origin possessing small values. In the inset of Fig. \ref{fig:Fig4b}, we show the phases (complex arguments) of the diagonal elements $[\widehat{T}^8]_{jj}$ as function of their own index $j$ and it is clear that all of them are very close to the common phase $\pi$, indicated by a dashed line.

\begin{figure}[ht!]
\centering
\subfigure[]{\includegraphics[width=3.9cm]{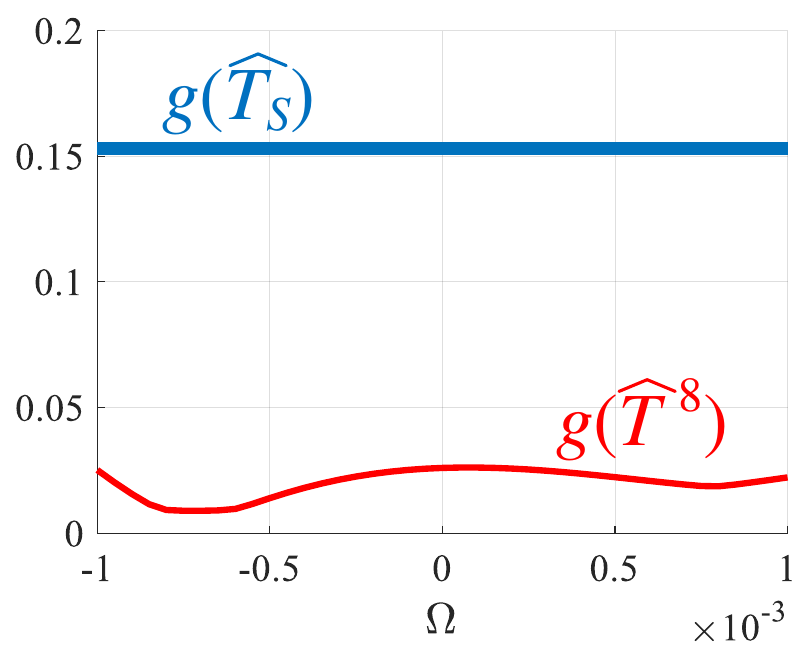}
   \label{fig:Fig5a}}
\subfigure[]{\includegraphics[width=4.4cm]{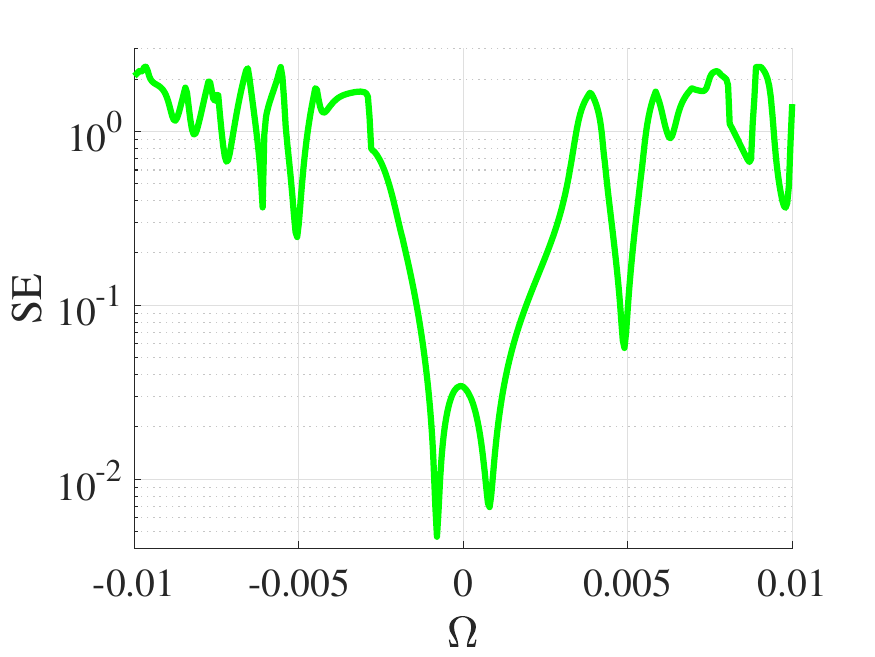}
   \label{fig:Fig5b}}
\caption{(a) The non-unitarity indicator $g$ of the (eighth power of) entire unit cell transfer matrix $\widehat{T}^8$ compared to the one of a single scatterer $\widehat{T}_S$ as function of the normalized frequency difference $\W=\frac{\omega-\omega_0}{\omega_0}$. (a) The standard error deviation {\rm SE} of the diagonal elements of the matrix $\widehat{T}^8$ as function of $\W$.}
\label{fig:Figs5}
\end{figure}

Given the fact that the information signal is not a single tone but distributes its power in other frequencies around $\omega=\omega_0$, it is meaningful to test the performance of our method in the vicinity of $\W=0$. In Fig. \ref{fig:Fig5a}, we represent the non-unitarity indicator $g$ of the overall matrix $\widehat{T}^8$ as a function of $\W$ and we assess that it is much smaller than the corresponding quantity of the single scatterer $g(\widehat{T}_S)\cong 0.15$ for a moderate range around $\w=\w_0$. Finally, in Fig. \ref{fig:Fig5b} we show the standard error ${\rm SE}$ deviation of the diagonal elements of the matrix $\widehat{T}^8$ from the ideal value $(-1)$ leading to dispersionless propagation, as a function of $\W$  across a more extensive band ($0.01<\W$<0.01). We notice its extremely small value retained throughout the part of the band used in Fig \ref{fig:Fig5a} ($0.001<\W<0.001$), which demonstrate{\color{blue}s} the success of the proposed and followed technique. On the other hand, we can observe the substantial (by more than two orders of magnitude) increase in ${\rm SE}$ for $|\W|>0.001$, which restricts the operational frequencies that dispersion-free transmission is achieved.

 With this example, we have demonstrated that our method can correct dispersion together with back-scattering effects up to second order; however, this also indicates some characteristics of the scattering model essential for a realistic implementation of the proposed method. In particular, one should choose (or eventually engineer) a scatterer without degeneracies on the eigenvalues of the involved generators providing simultaneously large mixing of angles at a weak back-scattering regime.

\section{Concluding Remarks \label{SV}}
The method that we describe in this work provides a pathway for coherent and dispersion-free propagation via suitable placement of weak scatterers that couple the supported modes. We report elimination of discrepancies in group velocities of different modes, vanishing quadratic dispersion at any Taylor order and  dispersion compensation at higher orders. The resolutions are applicable to both classical and quantum multimode networks, however we think that quantum communications, with their various multimode entangled states, offer more opportunities for applications of the suggested technique. Our approach can produce phase matching conditions along extended distances and thereby induce nonlinear coupling between the quantum fields of different modes, even for a typically weak nonlinear permeability. 

Even though the examples provided in this paper concern spatial transversal modes, the method is equally applicable to angular momentum vectors  which are currently attracting increasing interest \cite{AM} as well as to polarization states of quantum light. Finally, if seen from another point of view, we develop a method for achieving an effective ``unity'' in propagation. But starting from ``unity'' one can use numerical methods in order to engineer any desired unitary transform \cite{FE}; thus, our results can be simply extended to design arbitrary signal transformers by using as elements non-ideal scatterers placed at specified positions.

\section*{Acknowledgement}
All the authors express their gratitude to Sergey A. Moiseev (Kazan National Research Technical University, Russia) for numerous useful discussions and feedback on the draft. AM  acknowledges financial support from a Nazarbayev University ORAU grant ``Dissecting the collective dynamics of arrays of superconducting circuits and quantum metamaterials'' (no. SST2017031) and CV from Nazarbayev University Small grant ``Super transmitters, radiators and lenses via photonic synthetic matter'' (no. 090118FD5349). CV and AM also acknowledge MES RK state-targeted program BR05236454. VA thanks Grigory A. Kabatiansky (Skolkovo Institute of Science and Technology, Russia) for attracting his attention to this problem.  Finally, we would like to thank the third unknown referee of this work for essential feedback.

\end{document}